\documentclass[%
reprint,
 amsmath,amssymb,
 aps,
]{revtex4-1}

\usepackage{graphicx}
\usepackage{dcolumn}
\usepackage{bm}

\usepackage{cancel}
\usepackage{xspace}
\usepackage{lineno,hyperref}
\usepackage{xcolor}

\newcommand{\pt}{\ensuremath{p_{\rm T}}\xspace}

\newcommand{\py}{PYTHIA\xspace}

\begin{document}

\preprint{APS/123-QED}

\title{Energy dependence of underlying-event observables from RHIC to LHC energies}%
\author{Antonio Ortiz}
\email{antonio.ortiz@nucleares.unam.mx}
\affiliation{%
Instituto de Ciencias Nucleares, Universidad Nacional Aut\'onoma de M\'exico,\\
 Apartado Postal 70-543, M\'exico Distrito Federal 04510, M\'exico 
}

\date{\today}

\begin{abstract}
A study of the charged-particle density (number density) in the transverse region of the di-hadron correlations exploiting the existing pp and p$\bar{\rm p}$ data from RHIC to LHC energies is reported. 
This region has contributions from the Underlying Event (UE) as well as from Initial- and Final-State Radiation (ISR-FSR). Based on the data, a two-component model is built. This has the functional form $\propto s^{\alpha}+\beta\log(s)$, where the logarithmic (($\beta = 0.140 \pm 0.007$)) and the power-law ($\alpha = 0.270 \pm 0.005$) terms describe the components more sensitive to the ISR-FSR and UE contributions, respectively. The model describes the data from RHIC to LHC energies, the extrapolation to higher energies indicates that at around $\sqrt{s} \approx 100$\,TeV the number density associated to UE will match that from ISR-FSR. Although this behaviour is not predicted by PYTHIA~8.244, the power-law behaviour of the UE contribution is consistent with the energy dependence of the parameter that regulates Multiparton Interactions. Using simulations, KNO-like scaling properties of the multiplicity distributions in the regions sensitive to either UE or ISR-FSR are also discussed. The results presented  here can  be  helpful  to  constrain  QCD-inspired Monte  Carlo models at the Future Circular Collider energies, as well as to characterize the UE-based event classifiers which are currently used at the LHC.

\end{abstract}

\maketitle


\section{\label{sec:level1}Introduction}

The inelastic proton-proton (pp) cross section has contributions from diffractive (single diffraction and double-diffraction) and non-diffractive processes. For non-diffractive processes, occasionally, a hard parton-parton scattering occurs producing jets and high transverse momentum ($p_{\rm T}$) particles. The  Underlying Event (UE) consists of  particles from the proton break-up (beam-beam remnants) and the Multiparton Interactions (MPI) that  accompany such a hard scattering~\cite{Field:2012kd}. Multiparton Interactions, i.e. two or more semi-hard  parton-parton scatterings within the same pp collision, is a natural consequence given the composite nature of hadrons~\cite{Sjostrand:1987su}. Several data support the presence of MPI in hadronic interactions ~\cite{Akesson:1986iv,Abe:1997xk,Abazov:2009gc,Acosta:2004wqa,Aaltonen:2010rm,Adam:2019xpp,Aad:2010fh,ALICE:2011ac,Aad:2014hia,Abelev:2013sqa,Khachatryan:2015jza,Aaboud:2017fwp,Acharya:2019nqn,Ortiz:2020rwg,Ortiz:2021peu} The successful description of pp collisions by Monte Carlo (MC) generators relies on the precise modeling of UE~\cite{Sjostrand:1987su}. 

The study of pp collisions is also important given the discovery of heavy ion-like effects in high-multiplicity pp collisions~\cite{Nagle:2018nvi}: long-range azimuthal correlations~\cite{Khachatryan:2016txc}, radial flow~\cite{Acharya:2018orn} and strangeness enhancement~\cite{ALICE:2017jyt}. In heavy-ion collisions those effects are attributed to the production of a deconfined hot and dense QCD medium, known as the strongly-interacting Quark-Gluon Plasma~\cite{Bala:2016hlf,Busza:2018rrf}. Due to, e.g., MPI and color reconnection (CR) can produce collective-like effects~\cite{Ortiz:2013yxa}, the particle production as a function of quantities sensitive to MPI has attracted the interest of the heavy-ion community~\cite{Martin:2016igp,Ortiz:2018vgc,Zaccolo:2019hxt,Nassirpour:2749160,Ortiz:2020dph}.  

Experimentally, inelastic pp collisions are selected using a minimum-bias trigger, while the UE has to be studied in events in which a hard scattering has occurred.  This can be achieved selecting events with a high transverse momentum (e.g. $p_{\rm T}^{\rm trig.}\geq5$\,GeV/$c$) charged particle at mid-pseudorapidity. The activity in the transverse region of the di-hadron correlations is the most sensitive to UE, but it also has contributions from Initial- and Final-State Radiation (ISR-FSR).  In this paper, the available underlying-event data measured at the Relativistic Heavy Ion Collider (RHIC), the Tevatron, and the Large Hadron Collider (LHC) energies are investigated. The CDF Collaboration subdivided the transverse region into trans-max and trans-min in order to increase the sensitivity to ISR-FSR and UE (beam-beam remnant and MPI) effects, respectively~\cite{Aaltonen:2015aoa}. The UE component was found to increase like a power of the center-of-mass energy, while the ISR-FSR component increased logarithmically. Moreover, according with the MC generators PYTHIA~8.244~\cite{Sjostrand:2014zea} and HERWIG~7.2~\cite{HERWIGcollab_a2017a}, the \pt spectra and the particle composition are significantly different in the trans-max and trans-min regions~\cite{Bencedi:2021tst}. The present paper reports a data-driven model based on the UE measurements at RHIC, Tevatron and LHC energies. Last but not least, the physics opportunities at the Future Circular Collider (FCC) include the effects of the MPI mechanism and its connection with heavy-ion-like phenomena in proton-proton collisions~\cite{Mangano:2017tke,Benedikt:2018csr}. Therefore, the data-driven predictions are extended up to $\sqrt{s}\approx100$\,TeV. The results are compared with the \py~8.244 MC generator (Monash 2013 tune~\cite{Skands:2014pea}), hereinafter referred to as \py~8, for pp collisions at $\sqrt{s}$ from 0.2\,TeV up to 100\,TeV.  Finally, in order to improve the understanding on the scaling properties of the multiplicity distributions in the transverse region~\cite{Ortiz:2017jaz}, the studies are extended to higher multiplicities, and lower and higher center-of-mass energies and for the transverse, trans-max and trans-min regions.

The article is organised as follows:  Section II provides information about the analysis approach, as well as, the simulations using the \py~8 Monte Carlo generator. Section III presents the results and discussion, and finally section IV summarizes the results.

\section{\label{sec:analysis}Underlying Event observables}

The underlying-event analysis starts from the selection of the highest transverse momentum ($p_{\rm T}^{\rm trig.}$) charged particle of the event. The transverse region is defined by the associated particles within $\frac{\pi}{3} < |\Delta\phi| < \frac{2\pi}{3}$, where $\Delta\phi$ is the relative azimuthal  angle, $\Delta\phi = \phi^{\rm trig.}-\phi^{\rm assoc.}$, being $\phi^{\rm trig.}$ ($\phi^{\rm assoc.}$) the azimuthal angle of the trigger (associated) particle~\cite{Field:2012kd}. The charged-particle density in the transverse region (number density) is known to rise steeply for low values of $p_{\rm T}^{\rm trig.}$ and reaches a plateau at $p_{\rm T}^{\rm trig.}\approx 5$\,GeV/$c$ (see e.g. Ref.~\cite{Acharya:2019nqn}). This is due to the imposition of a hard trigger which biases the event selection towards events with many MPI which saturates at $p_{\rm T}^{\rm trig.}\approx 5$\,GeV/$c$. UE data include the average multiplicity at the plateau considering only primary charged particles with $p_{\rm T}\geq0.5$\,GeV/$c$. This allows for the comparison among different experiments.  

\begin{figure}[t]
\includegraphics[width=0.46\textwidth]{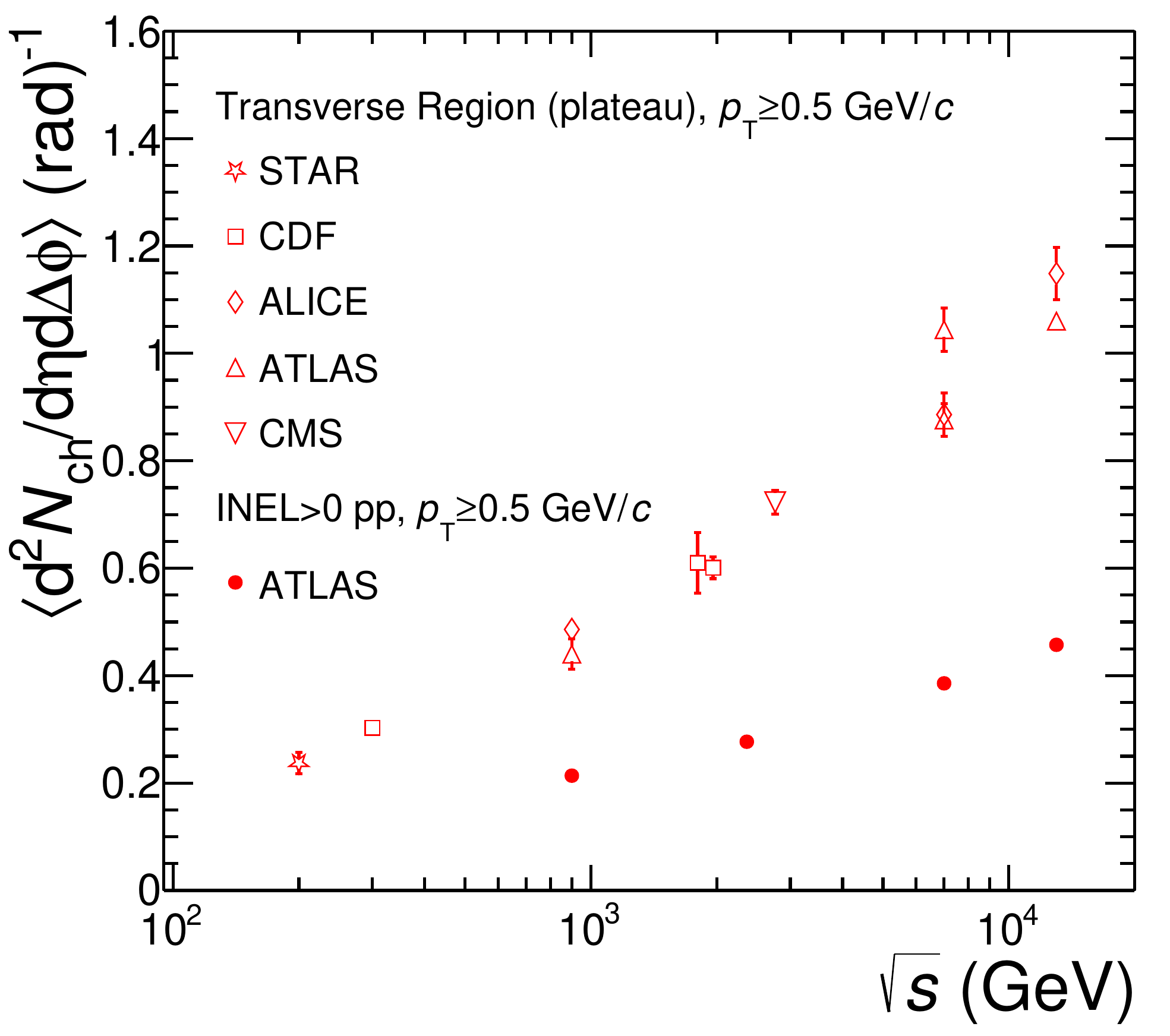}
\caption{Center-of-mass energy dependence of the number density reported by various experiments at RHIC, Tevatron and LHC. Except the ALICE data at 13 TeV, all the values were taken from Ref.~\cite{Adam:2019xpp}. These measurements are compared with $\langle {\rm d}N_{\rm ch}/{\rm d}\eta \rangle$ in INEL$>$0 pp collisions (scaled by 1/2$\pi$)~\cite{Aad:2010ac,Aad:2016mok}.  The number densities are obtained considering final state charged-particles with $p_{\rm T}\geq0.5$\,GeV/$c$. Error bars represent statistical and systematic uncertainties summed in quadrature.}
\label{fig1}  
\end{figure}

Figure~\ref{fig1} shows the number density for various experiments at RHIC~\cite{Adam:2019xpp}, Tevatron~\cite{Affolder:2001xt,Aaltonen:2015aoa} and LHC~\cite{Aad:2010fh,ALICE:2011ac,Aad:2014hia,Khachatryan:2015jza,Aaboud:2017fwp,Acharya:2019nqn}. Except the ALICE data point for pp collisions at 13\,TeV, all the values were taken from the compilation reported by the STAR Collaboration~\cite{Adam:2019xpp}. While the activity shows a modest increase from $\sqrt{s}=0.2$ up to 0.9\,TeV, for higher energies, it exhibits a steeper rise. The behaviour at higher energies is qualitatively similar to the center of mass energy dependence of the average number of MPI. The number density is compared with the average charged particle density (scaled by $2\pi$) measured by ATLAS~\cite{Aad:2010ac,Aad:2016mok} for inelastic pp collisions. As reported by the STAR Collaboration~\cite{Adam:2019xpp}, the number density increases faster with the center-of-mass energy than the average multiplicity in inelastic pp collisions. In order to investigate whether this behaviour is attributed to UE or ISR-FSR, a further treatment of the transverse side is implemented. 

The transverse region is sub-divided in two regions: 

\begin{itemize}
    \item transverse-I: $\pi/3<\Delta\phi<2\pi/3$
    \item transverse-II: $\pi/3<-\Delta\phi<2\pi/3$
\end{itemize}

The overall transverse region corresponds to combining the transverse-I and transverse-II regions. These two distinct regions are characterized in terms of their relative charged-particle multiplicities. Trans-max. (trans-min.) refers to the transverse region (I or II) with the largest (smallest) number of charged particles. According with earlier investigations~\cite{Aaltonen:2015aoa,Bencedi:2021tst}, these sub-regions  help to separate  the  ISR-FSR from the UE component of the collision. In the next section, the available trans-max and trans-min data from the CDF experiment will be used to build a model aimed at describing the activity in the transverse region. Results will be compared with \py~8. This MC generator is able to describe the underlying-event activity in a hard scattering process, and it was observed earlier that it well describes the measured data by several experiments at the LHC, see e.g.~\cite{Khachatryan:2010pv,Chatrchyan:2011id,Aad:2010fh,ALICE:2011ac}. The simulations consist of $5\times10^{8}$ inelastic pp collisions for each center-of-mass energy ($\sqrt{s}=0.2$, 0.9, 2.36, 7, 13, 50 and 100\,TeV). Only final state charged particles were accepted excluding the weak decays of strange particles in order to meet the experimental conditions. Only events with a trigger charged-particle with $p_{\rm T}^{\rm trig.} \geq 5$\,GeV/$c$ are considered.

\section{\label{sec:results}Results and Discussion}

\begin{figure}[t]
\includegraphics[width=0.46\textwidth]{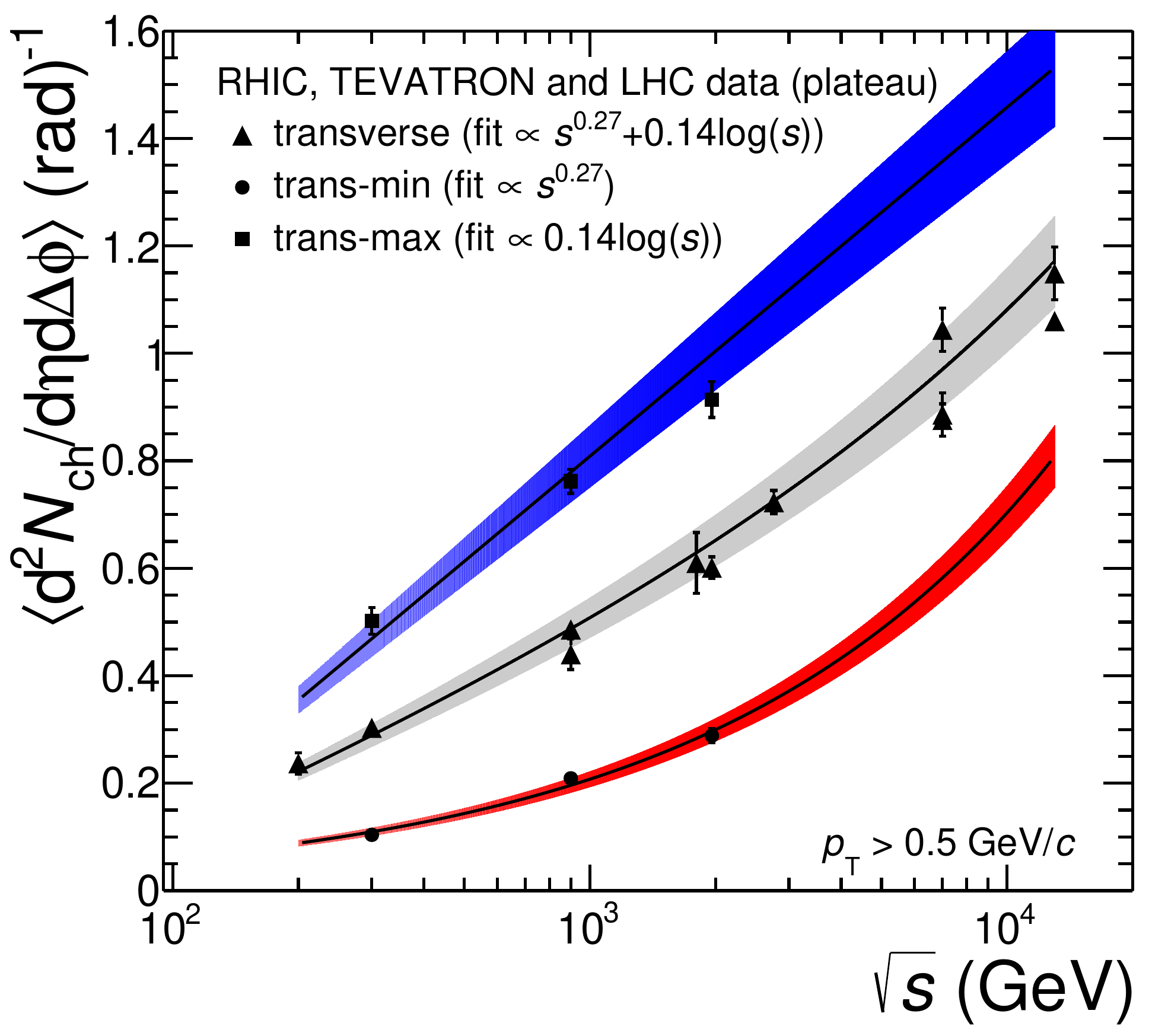}
\caption{Center-of-mass energy dependence of the average charged particle density  in the transverse, trans-min and trans-max regions at the plateau in pp~\cite{Adam:2019xpp,Affolder:2001xt,Aaltonen:2015aoa,Aad:2010fh,ALICE:2011ac,Aad:2014hia,Khachatryan:2015jza,Aaboud:2017fwp,Acharya:2019nqn} and/or p$\bar{\rm p}$ collisions~\cite{Aaltonen:2015aoa}. The number densities are obtained considering final state charged-particles with $p_{\rm T}>0.5$\,GeV/$c$. The data for the transverse region are compared with a parametrization of the form $s^{\alpha}+\beta \log(s)$, where the first term describes the MPI-sensitive region and the second one describes that more sensitive to ISR-FSR. The shaded areas indicate the one sigma systematic uncertainty.}
\label{fig2}  
\end{figure}

The CDF Collaboration has measured the number density in the transverse regions for p$\bar{\rm p}$ collisions from $\sqrt{s}=0.3$ up to 1.96\,TeV~\cite{Aaltonen:2015aoa}. To allow for comparisons with experiments at the LHC, the measurement considered charged particles within $p_{\rm T}\geq0.5$\,GeV/$c$ and $|\eta|<0.8$. The number density for trans-min (more sensitive to MPI) was found to increase much faster with the center-of-mass energy than does the trans-max (more sensitive to ISR-FSR). The CDF results are shown in Fig.~\ref{fig2} along with the data discussed in Fig.~\ref{fig1}. 

Motivated by the CDF observation that trans-dif, i.e. trans-max$-$trans-min, increases logarithmically with $\sqrt{s}$, while the trans-min increases like a power of the center-of-mass energy. A function of the form $s^{\alpha}+\beta \log(s)$ was simultaneously fitted to the existing data on number density in the transverse, trans-min and trans-max regions. Where the terms $s^{\alpha}$  and $\log(s)$ were constrained by the trans-min and trans-max data, respectively. The transverse region was found to be described by such a function with the parameters $\alpha = 0.270\pm 0.005$ and $\beta=0.140 \pm 0.007$. Given that in data the systematic uncertainty is significantly larger than the statistical one, the systematic errors from the data were propagated to the parametrization as follows. Using a random number generator, each data point was shifted up and down within one sigma of the systematic uncertainty. The two component model was then fitted to the data. The process was repeated 500 times.  The sigma of the distribution: (data-fit)/fit was assigned as systematic uncertainty to the parametrization. It amounts to around 7.2\%, and by construction, it is constant as a function of the center-of-mass energy. The determination of the systematic uncertainties of the parameters $\alpha$ and $\beta$ followed a similar procedure. The parametrizations along with the one sigma systematic uncertainty are shown in Fig.~\ref{fig2}. Within uncertainties, the number density as a function of $\sqrt{s}$ measured at RHIC, Tevatron and LHC energies are well described by the two-component data-driven model. It is worth mentioning that only published data are shown in the figure, however, the ATLAS Collaboration has preliminary results for trans-min and trans-max~\cite{Tokar:2017zid}. The preliminary number density for trans-min (trans-max) is around 0.82 (1.34) at $p_{\rm T}^{\rm trig.}=5$\,GeV/$c$ for pp collisions at $\sqrt{s}=13$\,TeV which is slightly above (below) the data-driven prediction.  Based on this parametrization, we observe that the activity in trans-min increases faster than trans-max. For example, at $\sqrt{s}=0.9$\,TeV the activity in trans-max (trans-min) is $\approx 0.78$ ($\approx 0.2$), whereas the activity at $\sqrt{s}=13$\,TeV is $\approx 1.53$ ($\approx 0.81$). This suggests that the MPI contribution increases by a factor $\approx 4$, while that which has a contribution from ISR-FSR increase by about a factor $\approx 2$. Regarding the charged particle density for inelastic pp collisions, the increase from $\sqrt{s}=0.9$\,TeV to 13\,TeV is slightly higher than 2.

\begin{figure}[t]
\includegraphics[width=0.46\textwidth]{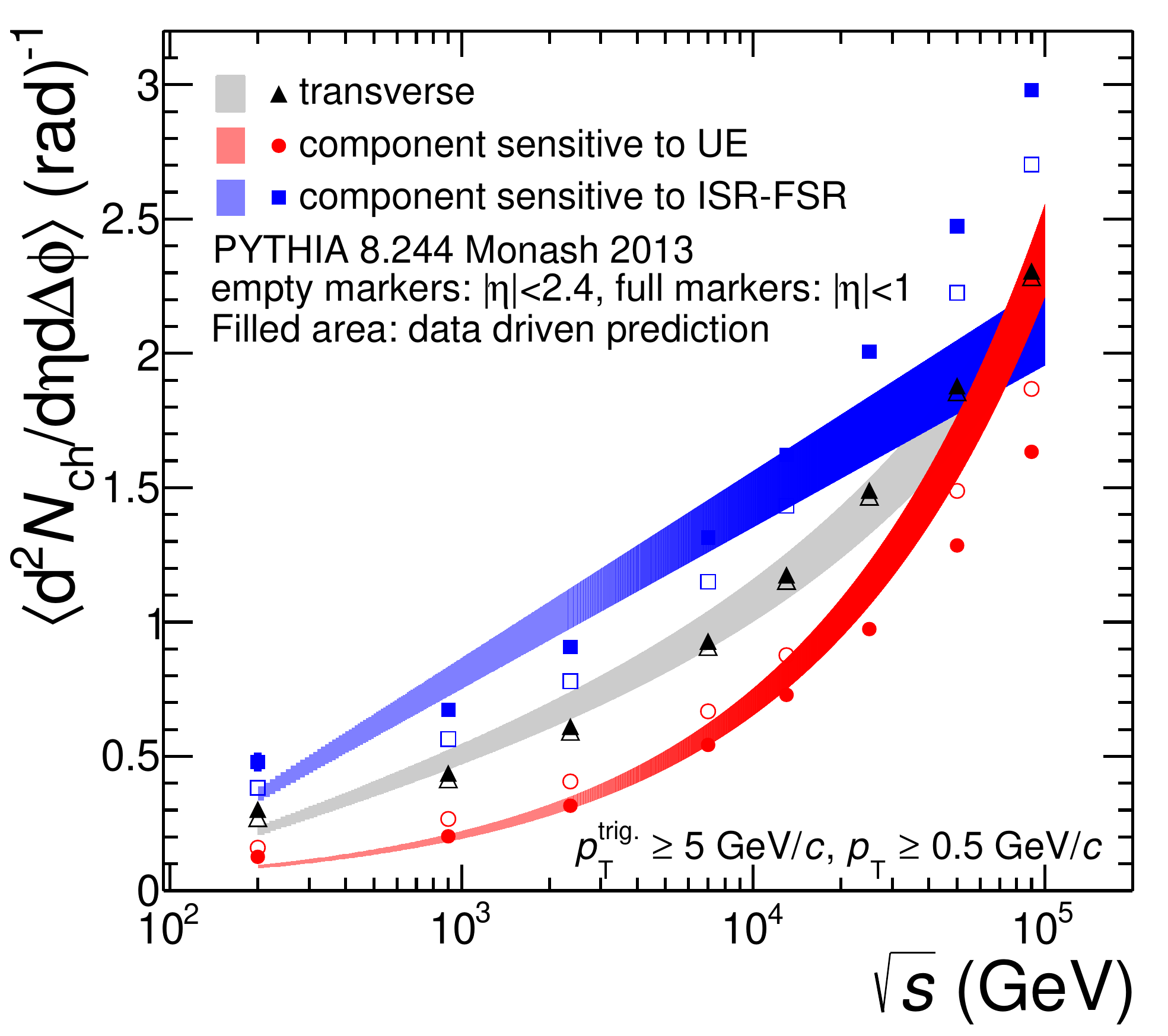}
\caption{Center-of-mass energy dependence of the average charged particle density in the MPI- and ISR-FSR-sensitive topological regions at the plateau ($p_{\rm T}^{\rm trig.}\geq 5$\,GeV/$c$) in pp collisions simulated with \py ~8 (tune Monash). Simulations are compared with the extrapolations using data at Tevatron energies (see the text for more details).}
\label{fig3}  
\end{figure}

\begin{figure*}
\begin{center}
\includegraphics[width=0.96\textwidth]{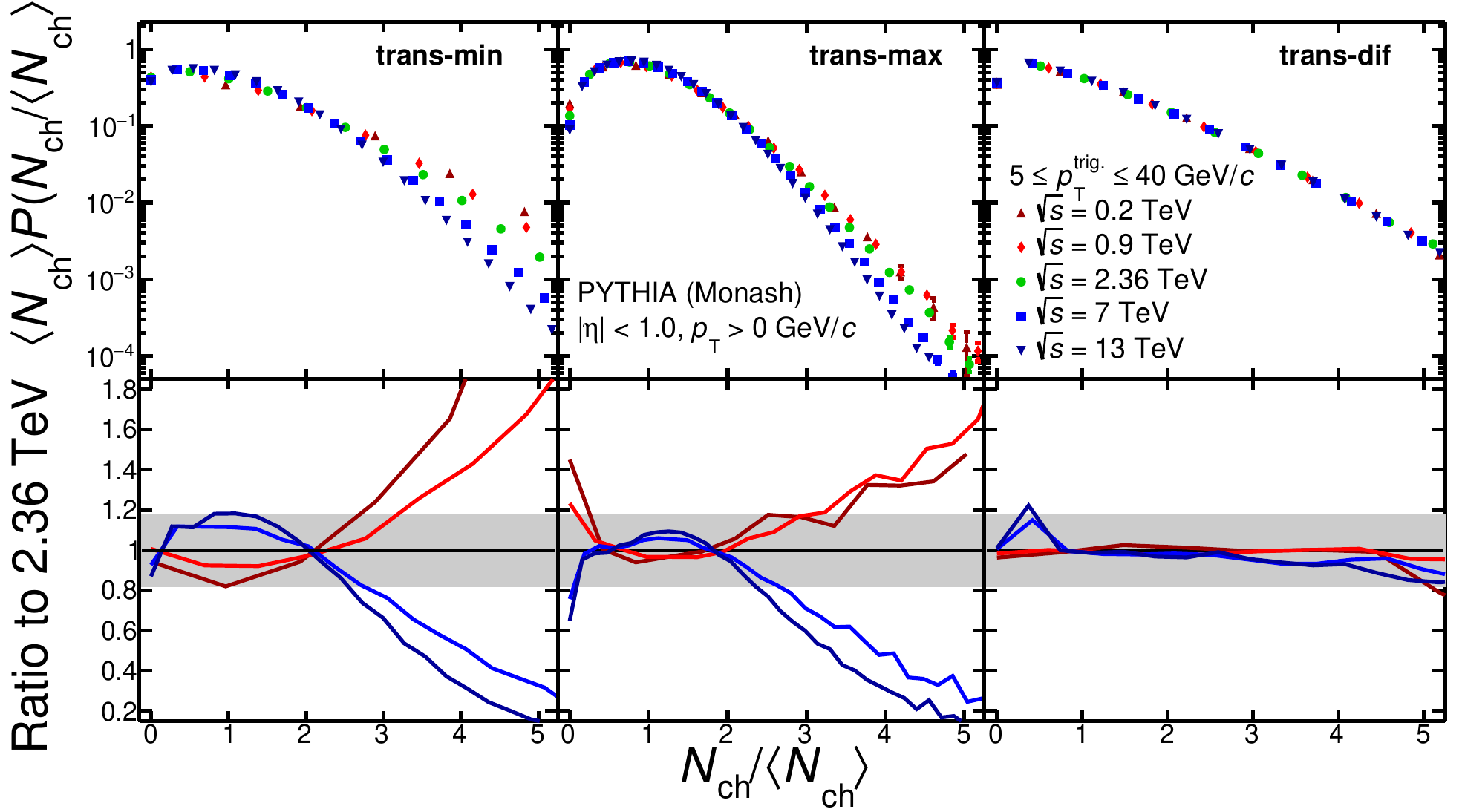}
\caption{Top: Charged particles multiplicity distributions in KNO variables in pp collisions simulated with \py~8. Results for different center-of-mass energies are shown for the trans-min (left), trans-max (middle) and trans-dif (right) regions. Bottom: The KNO multiplicity distributions are normalized to that for pp collisions at $\sqrt{s}=2.36$\,TeV.}
\label{fig4}  
\end{center}
\end{figure*}

In MC generators the energy evolution of MPI is implemented phenomenologically through a transverse momentum cutoff,  $p_{\rm T0}$, of a few GeV/$c$.  In  the  original PYTHIA modeling,  the  energy dependence of the total cross section was taken as the guideline for the energy evolution of $p_{\rm T 0}$~\cite{Sjostrand:1987su}. Namely, the perturbative  MPI  cross  sections  are  suppressed  below  the  $p_{\rm T0}$ scale, whose evolution with center-of-mass energy is driven by a power law:
\begin{equation}
   p_{\rm T 0}^{2}(s)=p_{\rm T 0}^{2}(s_0)\Big( \frac{s}{s_0}\Big)^b 
\end{equation}
Therefore, a higher scaling power $b$ implies a lower increase of the overall MPI activity. Modern tunes of PYTHIA yield $b$ values in the range $0.21-0.26$~\cite{Mangano:2017tke}. For example, the Monash tune considers $b=0.215$ and $\sqrt{s_0}=7$\,TeV~\cite{Skands:2014pea}. The number density as a function of $\sqrt{s}$ in \py~8 for the three topological regions are displayed in Fig.~\ref{fig3}. Results from $\sqrt{s}=0.2$\,TeV up to 100\,TeV considering charged particles within $|\eta|<1$ are displayed. A power-law function describes quite well the MPI-sensitive region (trans-min), the exponent is found to be $\approx 0.23$ which is below that obtained for data and close to the $b$ value which enters in the Monash tune~\cite{Skands:2014pea}.  Contrary to the data-driven prediction, a similar power-law behaviour is also observed for the ISR-FSR-sensitive region. In order to investigate the pseudorapidity dependence, \py~8 predictions are also displayed considering charged particles within $|\eta|<2.4$. While the transverse region is roughly pseudorapidity independent, the number density in trans-max (trans-min) exhibits an increase (decrease) with respect to the results considering a narrower pseudorapidity range. This suggests that hard radiation effects increases with the reduction of the pseudorapidity interval used in the analysis. Moreover, the number density increase from the lowest LHC energies ($\sqrt{s}=0.9$\,TeV) up to 100\,TeV is  around 4.4 (8.2) for the ISR-FSR-sensitive region (MPI-sensitive region). The data-driven model predicts that the increase of the activity in the MPI-sensitive region should be around 12 going from $\sqrt{s}=0.9$\,TeV up to 100\,TeV. Whereas, the increase for the ISR-FSR sensitive region is around 2.7 which is smaller than predicted by \py~8.  The discrepancy at FCC energies could be due to the tuning of soft
and semi-hard physics in Monte Carlo event generators which relies on Parton Distribution Functions (PDFs) in unexplored kinematical regions (e.g. $x \lesssim 10^{-5}$). Therefore, the data driven model could be useful to improve the MC predictions at FCC energies.

In the analysis of data, the relative contribution from MPI with respect to that from ISR-FSR is expected to play a role in observables like \pt spectra of unidentified charged particles, as well as, in particle ratios like $({\rm p}+\bar{\rm p})/(\pi^{+}+\pi^{-})$ as a function of \pt. If radiation plays an important role, then the particle ratios will be significantly suppressed with increasing the event multiplicity in the transverse side. On the other hand, if the MPI component is the dominant contribution, then the particle ratios will be significantly enhanced at intermediate \pt with increasing the event activity in the transverse side. This has been reported in Ref.~\cite{Bencedi:2021tst}, where the features of particle ratios as well as \pt spectra as a function of the activity in transverse, trans-max and trans-min regions were investigated. The preliminary ALICE data~\cite{Nassirpour:2749160} suggest that the relative contribution from ISR-FSR with respect to that from MPI is smaller in data than in \py~8~\cite{Bencedi:2021tst}. Early LHC data already suggested that the MPI activity at the LHC energies was already higher than in PYTHIA~\cite{Abelev:2012sk}. 

Last but not least, it has been reported that, within 20\%, the multiplicity distributions in the transverse region ($|\eta|<2.5$, $p_{\rm T}>0$\,GeV/$c$) at the plateau obey a Koba-Nielsen-Olesen (KNO) scaling~\cite{Ortiz:2017jaz}. This scaling was expected in a model which assumes that a single pp collision results from the superposition of a given number of elementary partonic collisions emitting independently~\cite{DiasdeDeus:1997ui}. Therefore, MPI could produce such an effect. In~\cite{Ortiz:2017jaz}, it was shown that the scaling held for pp collisions at the LHC energies for $0.5<z(=N_{\rm ch}/\langle N_{\rm ch}\rangle)<2.5$. Now, a refinement of that result is reported.  The prime goal is to investigate the KNO-like scaling properties if the sensitivity to MPI is improved. The impact of radiation is investigated using the multiplicity distributions in trans-max, as well as trans-dif. \py~8 results are reported for pp collisions from RHIC to LHC energies and for the pseudorapidity interval $|\eta|<1$ which allows to extend the $z$ reach. Figure~\ref{fig4} shows the multiplicity distributions in KNO variables for pp collisions at $\sqrt{s}=0.2$, 0.9, 2.36, 7, and 13\,TeV. The results for trans-max are qualitatively similar to those reported for the transverse region~\cite{Ortiz:2017jaz}. Namely, the KNO-like scaling holds for  $0.5<z<2.5$, whereas for lower or higher values of $z$ the violation of the KNO-like scaling is bigger than 20\%. It is worth noticing that for trans-max both contributions are considered: UE and ISR-FSR. If the effect of ISR-FSR is suppressed, i.e., exploiting the features of trans-min region, then the KNO-like scaling is extended up to very low multiplicities ($z<0.5$), whereas for $z>2.5$ the KNO-like scaling is still broken. Events with high multiplicity jets can contribute to the violation of the scaling properties. For example, a quantity sensitive to the number of MPI as a function of the event multiplicity is presented in Refs.~\cite{Abelev:2013sqa,Ortiz:2021peu}. It was observed that for $z>3$, the number of uncorrelated seeds (or MPI) deviate from the linear trend suggesting the presence of high multiplicity jets. Finally, results for trans-dif show a perfect KNO-scaling scaling in a broader $z$ interval, i.e. from 0 up to 5. This result complement the finding reported in Ref.~\cite{Ortiz:2017jaz}, suggesting that the hardest component of UE exhibits perfect scaling properties, whereas the MPI gives an approximate scaling which holds up to $z\approx2.5$. 

\section{Conclusions}

This article reports a data-driven model which is built using the existing UE data from RHIC up to LHC energies. The function which describes the data is of the form $\propto s^{0.27}+0.14\log(s)$, where the power-law term ($\alpha = 0.270 \pm 0.005$) and the logarithmic term ($\beta = 0.140 \pm 0.007$) describe the MPI- and ISR-FSR-sensitive topological region of the collision, respectively. Albeit in PYTHIA, the MPI-sensitive region is also well described by a power-law function ($\propto s^{0.23}$), such a contribution is found to increase faster with $\sqrt{s}$\, in the data-driven model than in PYTHIA~8.244. It is worth mentioning that the exponent which was found for \py~8 is close to that which enters in the Monash tune for the parametrization of the energy dependence of MPI. This paper also reports that at the FCC energies, the MPI-contribution is expected to dominate the transverse region, whereas an opposite behaviour is predicted by PYTHIA~8.244. One has to consider that the MC prediction relies on Parton Distribution Functions in the ultra low-$x$ regimen which has not been explored so far. Finally, the multiplicity distributions for each region were investigated considering pp collisions simulated with PYTHIA~8.244 from $\sqrt{s}=0.2$ up to 13\,TeV. A KNO-like scaling is predicted for the MPI-sensitive region, it would hold from $z=0$ up to $z=2.5$. For higher $z$ values, high multiplicity jets may break the scaling properties. The KNO-scaling is broken at low $z$ values ($z<0.5$) when radiation is folded together with the MPI contribution. However, when ISR-FSR effects are fully isolated, the scaling holds for a wide $z$ interval, from 0 up to 5. Data for trans-min and trans-max at the LHC energies would be needed in order to check the validity of the proposed parametrizations and the scaling properties.

\begin{acknowledgments}
The author acknowledges the very useful discussions with Guy Pai{\'c} and Peter Christiansen. Support for this work has been received from CONACyT under the Grant No. A1-S-22917.
\end{acknowledgments}

\bibliography{rt}

\end{document}